\documentclass[pmlr]{jmlr}%

\usepackage{longtable}%

\usepackage{booktabs}
\usepackage[load-configurations=version-1]{siunitx} %
\usepackage{svg}
\usepackage{float}
\definecolor{turquoise}{cmyk}{0.65,0,0.1,0.3}
\definecolor{purple}{rgb}{0.65,0,0.65}
\definecolor{dark_purple}{rgb}{0.45,0,0.45}
\definecolor{dark_green}{rgb}{0, 0.5, 0}
\definecolor{orange}{rgb}{0.8, 0.6, 0.2}
\definecolor{red}{rgb}{0.8, 0.2, 0.2}
\definecolor{darkred}{rgb}{0.6, 0.1, 0.05}
\definecolor{blueish}{rgb}{0.0, 0.3, .6}
\definecolor{light_gray}{rgb}{0.7, 0.7, .7}
\definecolor{pink}{rgb}{0.9, 0, 0.6}
\definecolor{greyblue}{rgb}{0.25, 0.25, 1}
\definecolor{teal}{rgb}{0.0, 0.4, 0.4}

\usepackage{subcaption}  %

\newcommand{\todo}[1]{{\color{red}#1}}

\newcommand{\ar}[1]{{\color{dark_green}#1}}

\newcommand{\da}[1]{{\color{magenta}#1}}

\newcommand{\ab}[1]{{\color{blue}#1}}

\def \customparskip {0.2em}
\renewcommand{\paragraph}[1]{\vspace{\customparskip}\noindent\textbf{#1}}

\renewcommand{\ar}[1]{#1}
\renewcommand{\da}[1]{#1}
\renewcommand{\todo}[1]{#1}
\renewcommand{\ab}[1]{#1}

\theorembodyfont{\upshape}
\theoremheaderfont{\scshape}
\theorempostheader{:}
\theoremsep{\newline}

\begin{document}

\begin{center}
    {\LARGE \bfseries Towards End-to-End Training of Automatic Speech Recognition for Nigerian Pidgin}\\
    [1em]
    \textit{Amina Mardiyyah Rufai\footnote{Equal contribution.}, 
    Afolabi Abeeb$^*$, 
    Esther Oduntan,
    Tayo Arulogun,
    Oluwabukola Adegboro\footnote{Co-supervisory role.},%
    Daniel Ajisafe$^\dagger$,
    }\\
    [1em]
    \texttt{[arufai, eoduntan, gadegboro, dajisafe]@aimsammi.org, aaabeeb@pgschool.lautech.edu.ng,
    otarulogun@lautech.edu.ng,
    }
\end{center}

\begin{abstract}
The prevalence of automatic speech recognition (ASR) systems in spoken language applications has increased significantly in recent years. Notably, many African languages lack sufficient linguistic resources to support the robustness of these systems. This paper focuses on the development of an end-to-end speech recognition system customized for Nigerian Pidgin English. We investigated and evaluated different pretrained state-of-the-art architectures on a new dataset. Our empirical results demonstrate a notable performance of the variant Wav2Vec2 XLSR-53 on our dataset, achieving a word error rate (WER) of 29.6\% on the test set, surpassing other architectures such as NEMO QUARTZNET and Wav2Vec2.0 BASE-100H in quantitative assessments. Additionally, we demonstrate that pretrained state-of-the-art architectures do not work well out-of-the-box. We performed zero-shot evaluation using XLSR-English as the baseline, chosen for its similarity to Nigerian Pidgin. This yielded a higher WER of 73.7\%. By adapting this architecture to nuances represented in our dataset, we reduce error by 59.84\%. Our dataset comprises 4,288 recorded utterances from 10 native speakers, partitioned into training, validation, and test sets. This study underscores the potential for improving ASR systems for under-resourced languages like Nigerian Pidgin English, contributing to greater inclusion in speech technology applications. We publicly release our unique parallel dataset (speech-to-text) on Nigerian Pidgin, as well as the model weights on Hugging Face. Our code \da{would be made} available 
to foster future research from the community. 

\end{abstract}

\section{Introduction}

Automatic Speech Recognition (ASR) plays a crucial role in transcribing spoken languages into text. Although ASR systems have become prevalent for widely spoken languages \citep{tamgno2012wolof}, the application of these systems to African languages remains a challenge. 
Africa, a continent known for its linguistic diversity with more than 2000 languages \citep{abbott2019benchmarking}, has a rich linguistic history shaped by language contact, expansion, trade language development, changes and instances of language dearth \citep{epstein1998language}. Most of these languages boast at least one million speakers each, contributing significantly to the global linguistic landscape \citep{epstein1998language}. Despite the linguistic richness, African languages face substantial challenges in terms of resources \citep{abbott2019benchmarking}. The majority of these languages are low-resourced \citep{laleye2016first}, hindering the development and robustness of ASR systems, especially when compared to more extensively resourced Western languages like English, French, and German.
This study focuses on `Pidgin English', one of the most widely spoken languages in Africa, with an estimated 75 million speakers in Nigeria and 5 million in Ghana. Although there are variations of this language, our research concentrates on the ``Nigerian Pidgin English", the most prevalent form in West Africa \citep{ogueji2019pidginunmt}.

\da{Prior work} created the first monolingual Pidgin text-to-text corpus and trained the first word vectors on this language~\citep{ogueji2019pidginunmt}. \da{This} involved aligning these vectors with English word vectors to produce cross-lingual embeddings. \da{Subsequently, improvements made} by using the English text as a pivot language in the target domain, improving the fluency and relevance of the Pidgin text~\citep{chang2020unsupervised}. 

Under the speech-to-text category, \cite{blachon2016parallel} began the first effort to develop human language technology (HLT) tools, specifically speech resources for Nigerian Pidgin. Their work focused on developing a speech corpus for a tokenizer, an automatic speech system for predicting the pronunciation of words and their segmentation. However, these resources were only integrated into a software tool, and we are not aware of any discoverable platform where the data has been made readily available for research purposes. 

In an effort to strengthen the Nigerian Pidgin language and its area of research, we \da{compiled} the first publicly available speech-to-text dataset on Nigerian Pidgin. We fine-tuned variants of the Wac2Vec2.0 architecture on this custom dataset. We compared the outcomes of these models against \da{the baseline} and achieved a lower error rate of 29.6\% on the test set. Our key contributions are; 
\begin{enumerate}
  \item To provide a publicly accessible (end-to-end) ASR system for Nigerian Pidgin
  \item Provide a free speech corpus for Nigerian Pidgin, and
  \item Present the first parallel (speech-to-text) data on the language as a benchmark for further research.

\end{enumerate}

The following sections are structured into the methodology, results, \da{ discussion}, and conclusion section. 
The methodology \da{section} details the use of the LIG-Aikuma App for linguistic documentation\da{,} 
data sources
and speech recording with \da{corresponding} preprocessing steps.
Under the Model Architecture section, 
we explained the evaluated models - Nemo quartznet, Wav2Vec 2.0 base-100H, and Wav2Vec xlsr-large-53 respectively - describing the design, training, and adaptations. The results section
presents the metrics used and compares architecture performances. Finally, the conclusion
summarizes the findings, emphasizes contributions, and suggests future research directions for ASR in low-resource languages.

\section{Methodology}
\label{methodology}

Developing an automatic speech recognition system requires an
adequate amount of speech recordings and corresponding text data. However, curating these \da{paired} resources is challenging, especially for low-resource languages. This is because they
are not readily or publicly available online. In this section, we describe the methodology used to collect speech recordings and textual data for building our automatic recognition system.

\subsection{Textual Data}
\label{textual_data}

 The traditional method for amassing substantial textual data for 
 ASR systems involve sourcing texts from online platforms. While many languages benefit from readily available resources like Wikipedia corpora, Pidgin English presents a different scenario. Though widely spoken by 
 millions of West Africans, \da{it lacks} adequate NLP resources \citep{ogueji2019pidginunmt}. \da{We build on prior data efforts made on the text-to-text parallel corpus, which was crawled from various news websites. The total crawled data comprised 56,695 sentences and 32,925 distinct words, covering topics ranging from sports, politics, and entertainment to everyday life.} From this, we selected a \ar{subset} of 4,288 utterances for recording speech data, \ar{with each utterance averaging between 8 and 17 words. The average sentence length in the corpus is 86 characters, with a corresponding mean audio duration of approximately 17 seconds.}

 \label{topic_model}
 \begin{figure}[htbp]
     \centering
     \includegraphics[width=0.9\textwidth]{fig/topic_distribution_2.pdf}
     \caption[]{This shows the topic distribution in the \da{Nigerian Pidgin} text dataset, revealing dominant themes such as “General/Everyday conversations” \da{(24.4\%), “Government/Politics” (18.6\%), and “Sports” (14.1\%), with \da{less-dominant themes} in areas such as “Telecommunication” (1.3\%) and “Agriculture (1\%).”}}
     \label{fig:topic_distro}

 \end{figure}

\paragraph{Topic distribution:}
\da{To understand the most dominant themes in the dataset, we performed a multi-stage unsupervised topic modelling analysis using BERTopic\citep{grootendorst2022bertopic}. We start with an initial clustering, which was subsequently improved using KeyBert\citep{grootendorst2020keybert}, and Maximal Marginal Relevance (MMR). This reduced topic outliers and enhanced coherence. By re-assigning and merging similar topics into one, we
come down to 15 distinct themes, where “General/Everyday Conversation,” and “Government/Politics” emerged as dominant themes, as shown in Figure~\ref{fig:topic_distro}.
}

\subsection{Speech Corpus}
\label{speech_corpus}

With the acquired text information, we proceed to collecting
speech recordings to build the recognition system. Given the absence of an existing speech corpus, we carried out the task of
recording and collecting our own speech data. {We used the LIG-Aikuma App for recording \citep{gauthier2016lig}.} 
The software made extensive data gathering possible due to its interoperability with Android devices and easy-to-share feature. In total, we amassed 4,288 instances of speech data, captured from 10 native pidgin speakers (comprising of 5 males and 5 females). Their ages ranged between 20 and 28 years,  and the data was captured in an environment with mostly minimal background noise, as much as possible.

\subsubsection{Speech Data Preprocessing}
\label{speech_preprocess}

Efficiently preprocessing speech data is crucial in developing robust speech recognition systems \citep{keerio2009preprocessing}. Therefore, our approach employs several key steps.

\paragraph{Audio Segmentation:}  To align with the requirements of the model architectures used in our experiments, we ensured that each audio file had a duration of no more than 30 seconds per chunk. Most of the audio files ranged between 1 and 17 seconds, with shorter audio clips being typical in this dataset. To further standardize the data, all audio files were resampled to a 16 kHz sampling rate, ensuring consistency across the dataset and compatibility with our processing pipeline. 

Additionally, corrupt files such as empty recordings, unintelligible speech, or audio containing only background noise were manually identified and removed by the data validator. These issues likely resulted from initial sentence boundary segmentation errors.
By resampling and excluding artifacts,
our preprocessing pipeline ensured high-quality, consistent inputs, reducing the risk of errors during model training, while improving overall data integrity. \ar{After data preprocessing and filtering, \da{the result is a final size of} 4,277 speech recordings subsequently partitioned into training, validation, and testing sets.} 

\paragraph{Feature Extraction:} While core principles like resampling (16 kHz) and normalisation are common across architectures, specific feature extraction steps differ depending on the model. For Nemo Quartznet,  we applied and used Mel Spectrogram \citep{kriman2020quartznet} as input rather than a linear input, to extract better feature representation such that differences in-between frequencies are more aligned to what humans perceive. In contrast, the Wav2vec2 models can handle raw waveforms directly $-$ replacing traditional preprocessing steps with a convolutional feature encoder.

\paragraph{Data Augmentation:} 
For Nemo Quartznet, we utilized SpecAugment, a well-established technique for automatic speech recognition \citep{park2019specaugment} and applied this to the feature inputs. For the Wav2Vec2 variants, we followed the 
augmentation protocols provided in \citep{patrickvonplaten}  
    
\subsection{MODEL ARCHITECTURES}
\label{model_archs}
We evaluated our data set on three (3) state-of-the-art architectures. We briefly discuss key components of these architectures below:

\subsubsection{Nemo Quartznet}
\label{nemo_model}
The Nemo QuartzNet, as detailed in \citep{kriman2020quartznet}, is based on a convolutional neural network framework and is trained using the CTC loss function \citep{graves2006connectionist}
Notably, this model exhibits resemblances to its forerunner, the Nemo Jasper architecture \citep{li2019jasper}. A notable deviation lies in the adoption of 1D time-channel separable convolutions, diverging from the conventional 1D convolution employed in Jasper. 
The model was fine-tuned for $30$ epochs using a greedy CTC decoder. We use NovoGrad as the optimizer with the same learning rate of $1e-4$ and weight decay of
$1e-3$.

\subsubsection{Wav2vec2.0 Base-100h}
\label{wav2vec_base}

Wav2Vec 2.0 \citep{baevski2020wav2vec} is an effective framework for learning powerful representations from speech data.
Its self-supervised training methodology allows for seamless model pre-training on unlabeled data, which subsequently can be fine-tuned on small labeled data as a downstream task. 
We unfreeze the feature encoder and finetune Wav2Vec 2.0 on Nigerian Pidgin to refine the model's ability to predict specific words or phonemes particular to the language.
The model was trained for 30 epochs but with an AdamW optimizer. The learning rate is set to $1e-4$.

\subsubsection{ Wav2vec xlsr-large-53}
\label{wav2vec_xlsr}

The XLSR is built on the wav2vec 2.0 architecture, which learns cross-lingual speech representations from multiple languages.
The model has a simultaneous learning process of quantized latent speech representations shared across languages \citep{conneau2020unsupervised}. These shared representations provide a strong generalization capability for unseen languages. 
We use the same implementation setting as the base wav2vec model.

\section{Results and discussion}
\label{results}
The word error rate (WER) is a key metric in automated speech recognition (ASR). It measures discrepancies between recognized and reference word sequences, as these sequences can sometimes be complicated by variable lengths. 
The word error rate can be computed as,

\begin{equation}
WER = \frac{S + D + I}{N} = \frac{S + D + I}{S + D + C}
\end{equation}

Where $S$ is the number of substitutions; $D$ is the number of deletions; $I$ is the number of insertions; $C$ is the number of correct words; and $N$ is the number of words in the reference $(N=S+D+C)$.

The model with the lowest WER in the unseen test set is considered the best-performing model. Quantitatively, Wav2Vec-XLSR53 demonstrated the best WER on the unseen test set, highlighting its effectiveness in handling the nuances and variations of the language, courtesy of its robust representations. Table \ref{tab:empirical_results} below presents the full quantitative results, with Nemo baseline achieving the lowest WER at a score of $0.566$ signaling the limitation of fully-supervised methods to cross-domain generalization task. We also show qualitative results for the XLSR model in Table \ref{tab:qualititative_results}. Our method is able to capture nuances specific to the target language such as ``sabi” (to understand), ``tok” (talk), ``pipo” (people) etc. $-$ signaling a reasonable generalization ability. However, we noticed that it sometimes struggles with numbering e.g. ``217” instead of ``27”.

\begin{table}[htbp]
\centering
\caption{This shows the quantitative results across all architectures used in this study. Highest scores are indicated in \textbf{bold}. Notably, wav2vec xlsr53 achieved the lowest WER on validation and test sets.}
\label{tab:empirical_results}

\begin{tabular}{lcc}
\toprule
Model & Val WER & Test WER \\
\midrule
XLSR-53-English (Baseline)\footnotemark{} &  -- & 0.737 \\
\midrule
Nemo & 0.547 & 0.566 \\
Wav2VecBase & 0.397 & 0.373 \\
\da{Wav2Vec XLSR-53 (Ours)} & \textbf{0.316} & \textbf{0.296} \\
\bottomrule
\end{tabular}
\end{table}

\begin{table}[H]
\caption{Qualitative Results for \da{Wav2Vec XLSR-53, our best performing model}. These demonstrate the model's ability to transcribe Nigerian Pidgin speech, highlighting its effectiveness in capturing language nuances.}
\label{tab:qualititative_results}
\centering
\small
\begin{tabular}{p{0.3\textwidth} p{0.3\textwidth} p{0.3\textwidth}}
\toprule
\bfseries Reference & \bfseries Our Prediction & \bfseries Zero-Shot Prediction \\
\midrule
\textbullet{} pipo and all di poor pipo wey govment gats take care of & pipo and all di poor pipo wey govrment gats take care of & people and ol the poor peopleway government gats take care of \\
\textbullet{} so dat one con mean say no show for dem next year & so dat one con mean say no show for dem next year & so thats on't calm me in senushu for them next year \\
\textbullet{} i no see why we no get proper health insurance & i no see why we no get proper health insurance & i' non's the why we know gepepiorts ensurance \\
\textbullet{} di social workers want make him pay her bill so dem & di social workers want make him pay habi so dem & in social wockarse one making paya bisodan \\
\bottomrule
\end{tabular}
\end{table}

\begin{table}[H]

\caption{Failure case for Wav2Vec2.0 XLSR-53. This highlights a limitation in transcribing Nigerian Pidgin speech, showing how numeric elements may be misinterpreted, thus affecting transcription accuracy.}
\label{tab:qualititative_limitation}
\centering
\resizebox{\textwidth}{!}{%
\begin{tabular}{ll}
\toprule
\textbf{Reference} & \textbf{Prediction} \\
\midrule
dem go wan kill chief femi fani kayode wife 27 & dem go wan kill chief femi fani kayode wife 217 \\
\bottomrule
\end{tabular}%
}
\end{table}

\footnotetext{\ab{Validation WER for XLSR-53-English (Baseline) is not reported, as this model was evaluated in a zero-shot setting on the test set only, without fine-tuning.}}

The quantitative results from our study on Automatic Speech Recognition (ASR) systems for Nigerian Pidgin English reveal significant implications for both research and practical applications. The evaluation metric used, WER, demonstrates that Wav2Vec XLSR-53 surpassed other models in accuracy and performance on both the validation and the test set. We hypothesize the key implications and reasons for this superior performance to the following:

{\begin{enumerate}
\item \textbf{Robustness and Adaptability:} Wav2Vec XLSR-53, built on the Wav2Vec 2.0 architecture, incorporates cross-lingual speech representations (XLSR), allowing it to learn latent speech features that are shared across multiple languages. This approach enhances the model's ability to handle variations and nuances specific to Nigerian Pidgin English, despite the language's under-resourced status.

\item \textbf{Fine-tuning Capability:} The model's self-supervised pre-training methodology followed by fine-tuning on Nigerian Pidgin English speech data plays a crucial role. The \todo{latter process refines} the model's ability to predict specific phonemes and words relevant to the target language, resulting in improved accuracy compared to baseline models such as Nemo and earlier versions of Wav2Vec.

\item \textbf{Dataset Suitability:} Our study utilized a comprehensive dataset comprising 4,288 instances of speech data from native Nigerian Pidgin speakers. This dataset, collected using the LIG-Aikuma app and augmented (during training) for robustness, provided ample training examples crucial for optimizing Wav2Vec XLSR-53's performance.
\end{enumerate}

The superior performance of Wav2Vec XLSR-53 in our study not only validates its efficacy but also highlights the potential of advanced ASR technologies in bridging the gap for under-resourced languages. This research contributes to the broader goal of inclusivity in artificial intelligence applications, particularly in diverse linguistic contexts such as those found in Africa.}

\section{Ethics and Limitation}

\ar{Our data collection process adhered to ethical standards, including informed consent and speaker privacy. 
While valuable as an initial resource, the dataset remains limited in size and speaker diversity, and it does not fully capture the dialectal variations across regions $-$ limiting generalization and robustness of ASR models trained on this resource. \da{Increasing these variations would be beneficial for cross-domain performance.}
}

\section{Conclusion}
\label{conclusion}

In this study, we introduced parallel speech-to-text data for Nigerian Pidgin as a benchmark to support accessible and reproducible research in automatic speech recognition (ASR) systems. Using a zero-shot approach, the pretrained model produced a WER of 73.7\%, reflecting the challenge of recognising Nigerian Pidgin without task-specific training. However, after fine-tuning on our curated dataset, the WER was substantially reduced to 29.6\%, marking a significant advancement in ASR performance for this under-resourced language. This improvement demonstrates the value of domain-specific data and adaptation strategies in building effective speech technologies for African languages.

Future work should focus on enhancing the quality and diversity of the training and evaluation dataset. Broader demographic and geographic representation in speech samples can improve model generalisation, and techniques such as unsupervised test-time speaker adaptation may further reduce error rates. 

Continued collaborative research is essential to ensure that languages like Nigerian Pidgin are well-represented in the digital sphere, fostering inclusive and equitable language technologies.

\bibliography{manuscript}  

\end{document}